# Reversible Wavefront Shaping Between Gaussian and Airy Beams By Mimicking Gravitational Field

Xiangyang Wang[1], Hui Liu[1, a], Chong Sheng[1] and Shining Zhu[1]

[1]*National Laboratory of Solid State Microstructures and School of Physics, Collaborative Innovation Center of Advanced Microstructures, Nanjing University, Nanjing, Jiangsu 210093, China*

[a]Corresponding author: e-mail: *liuhui@nju.edu.cn* .

## Abstract

In this paper, we experimentally demonstrate reversible wavefront shaping through mimicking gravitational field. A gradient-index micro-structured optical waveguide with special refractive index profile was constructed whose effective index satisfying a gravitational field profile. Inside the waveguide, an incident broad Gaussian beam is firstly transformed into an accelerating beam, and the generated accelerating beam is gradually changed back to a Gaussian beam afterwards. To validate our experiment, we performed full-wave continuum simulations that agree with the experimental results. Furthermore, a theoretical model was established to describe the evolution of the laser beam based on Landau's method, showing that the accelerating beam behaves like the Airy beam in the small range in which the linear potential approaches zero. To our knowledge, such a reversible wavefront shaping technique has not been reported before.

## 1. Introduction

The manipulation of the wavefront of light in different artificial systems, such as photonic crystals, surface plasmons and metasurfaces, has attracted increasing attention from researchers. With wavefront shaping, tremendous progress in refraction, illumination, optical imaging, and a host of other applications has been made. Up to now, how to better control wavefront shaping has become an increasingly important issue with the demand for new technologies. In 2006, Pendry *et al.* and Leonhardt proposed a transformation optics (TO) approach [1, 2] based on the invariance of Maxwell's equation under a coordinate transformation. In the last 10 years, with advances in materials science, many TO devices have been designed and developed to control electromagnetic waves, such as invisibility cloaks[3-8], filed rotators [9, 10], illusion optics [11], photonic black holes [12-15], Einstein rings [16], Mikaelian lenses [17], and nanofocusing plasmonics[18-20]. At the same time, research into accelerating beams, the wavefronts of which are designed to lead light beams capable of propagating along curved trajectories in free space and in a diffraction-free manner, has surged. These peculiar beams were first theoretically proposed in the framework of quantum mechanics in 1979 [21] and introduced into the optical domain in 2007 [22, 23]. Since then, many potential applications of such beams have been demonstrated, including imaging and microscopy [24], light bullets [25, 26], optical manipulation of microparticles [27], optical routing [28], plasmonic Airy beams [29-31], the Airy-Talbot effect [32], and analogs of gravitational effects [33]. However, these beams are not reversible; that is, one can transform Gaussian beams or planar waves to accelerating beams, but one cannot transform the accelerating beams back into Gaussian beams.

Here, we experimentally demonstrate reversible wavefront shaping between Gaussian beams and accelerating beams by mimicking gravitational fields in TO. In our experiment, an incident broad Gaussian beam is transformed into an accelerating beam in a gradient-index micro-structured optical waveguide with a special refractive index profile mimicking a gravitational field, and then the generated accelerating beam is gradually transformed back into a Gaussian beam. We perform full-wave continuum simulations using the beam-propagation method (BPM) [34] to validate our experimental findings. Based on Landau's method [35], we show that the accelerating beam is similar to an Airy beam in the small range in which the linear potential approaches zero. The reported TO waveguide can provide a technique for achieving reversible wavefront shaping that has not, to our knowledge, been previously reported.

## 2. Sample fabrication and measurement

A structured waveguide was fabricated as depicted in figure 1(a) (side view of the waveguide). First, a 50-nm-thick silver film was sputtered onto a silica substrate. Second, a straight silver cylinder was glued to the silver film, the silver cylinder was used to fabricate curved waveguide because of surface tension effects, and then a coupling grating with a period of 310 nm was milled on the silver film using a focused ion beam (FEI Strata FIB 201, 30 keV, 11 pA). Next, a polymethylmethacrylate (PMMA) resist mixed with $Eu^{3+}$ rare-earth ions, added for the purpose of fluorescence imaging, was deposited on the silver film using a spin-coating process. Finally, the sample was dried in an oven at 70°C for 2 h. During this fabrication process, the thickness of the PMMA layer can be controlled by varying the solubility of the PMMA solution, evaporation rate, and spin rate. As a result, an optical waveguide with continuously and smoothly variable-thickness is produced. In addition, the thickness of the waveguide remains unchanged in the direction parallel to the axis of the cylinder $z$ direction in figure 1(b), while in the perpendicular direction it shows first a slow increase and then a fast decrease along the $x$ direction in figure 1(a) and

1(b).

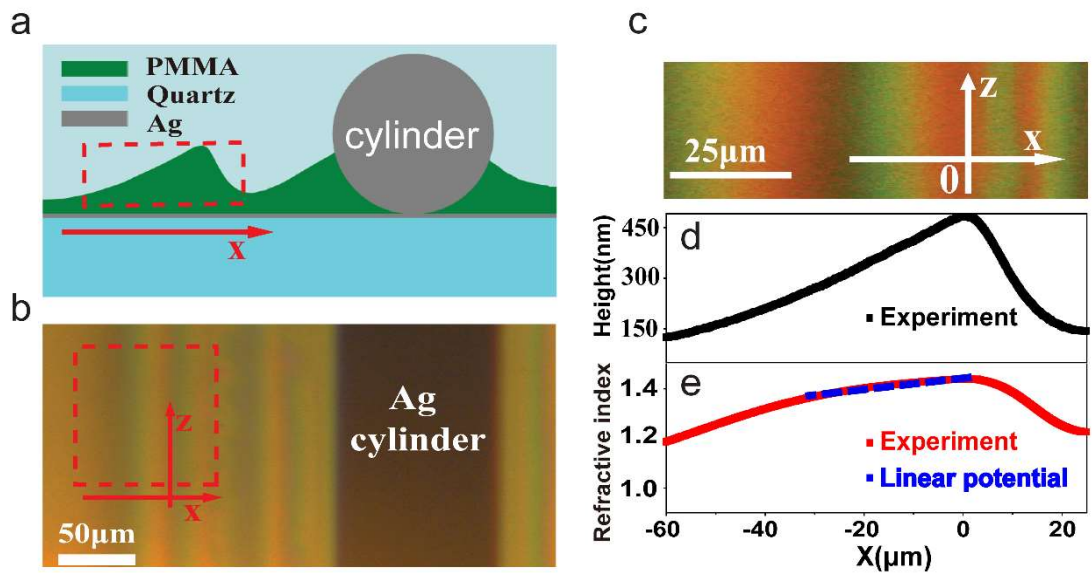

**Figure 1.** Schematic and structural measurements of the sample. (a) Schematic (side view) of the micro-structured optical waveguide. The waveguide is formed around a microcylinder with a diameter of 100 μm. (b) Microscopy image (top view) of the sample. The sample is illuminated by white light from the substrate, and the interference pattern arises only in the direction ($x$ direction) perpendicular to the axis of the cylinder ($z$ direction). (c) Interference pattern of the waveguide, enlarged by the red dashed-line box in (a) or (b). The $x$ direction of the coordinates is perpendicular to the axis of the cylinder, while the $z$ direction, the propagation direction, parallels the axis. (d) Surface profile of the PMMA layer measured with the step profiler along the $x$ direction. (e) Effective refractive index of the micro-structured waveguide, as shown by red solid line. Blue dash line shows the fitting results-a linear potential to mimic a uniform gravitational field based on TO considerations.

This curved surface of the waveguide can be indirectly observed by examining the interference pattern around the Ag cylinder shown in figure 1(b), which is a microscopy image (top view) of the sample illuminated by white light. With a higher-magnification objective lens, a clear interference pattern obtained from the area in the red dashed-line box in figure 1(a) or 1(b) is shown in figure 1(c). The analysis presented in the following sections is based on the coordinate system established in figure 1(c). Along the $x$ direction, we can see the clear colored interference stripes caused by the PMMA's graded thickness, which become closer and closer. Along the positive $z$ direction, however, no stripes can be observed. To directly obtain the surface profile of the PMMA layer in the red dashed-line box, we quantificationally measure the thickness profile (Fig. 1(d)) using a step profiler (Dektak® 150, Veeco Instruments, Inc., USA), and find that the surface profile corresponding to the red dashed-line box in figures 1(a) and 1(b) is in good agreement with the interference pattern in figure 1(c). Specifically, the waveguide thickness increases for $x<0$ and decreases for $x>0$.

According to previous works [8, 15-17, 36], the refractive index required by TO devices can be designed by tapered waveguides, and the light wave propagating in a variable-thickness waveguide can be described using the waveguide effective refractive index. In the experiment, the structured waveguide consists of an air/PMMA/silver/SiO2 multilayer stack (figure 1(a)) and can be considered a step-index planar waveguide. The dispersion relationship of the waveguide transverse electric (TE) modes is used to extract the effective refractive index in the $x$ direction, which is depicted by red solid line in figure 1(e). One can easily find that the waveguide effective refractive index $n_{\text{eff}}$ (for the $\text{TE}_0$ mode) increases with the $x$ direction for $x \leq 0$ and shows a linear dependence $n_{\text{eff}}^2 = n_{\text{eff}}^2(x) = ax + b$ in a small range, where the parameters $a$ and $b$ are constants retrieved from the fitting result.

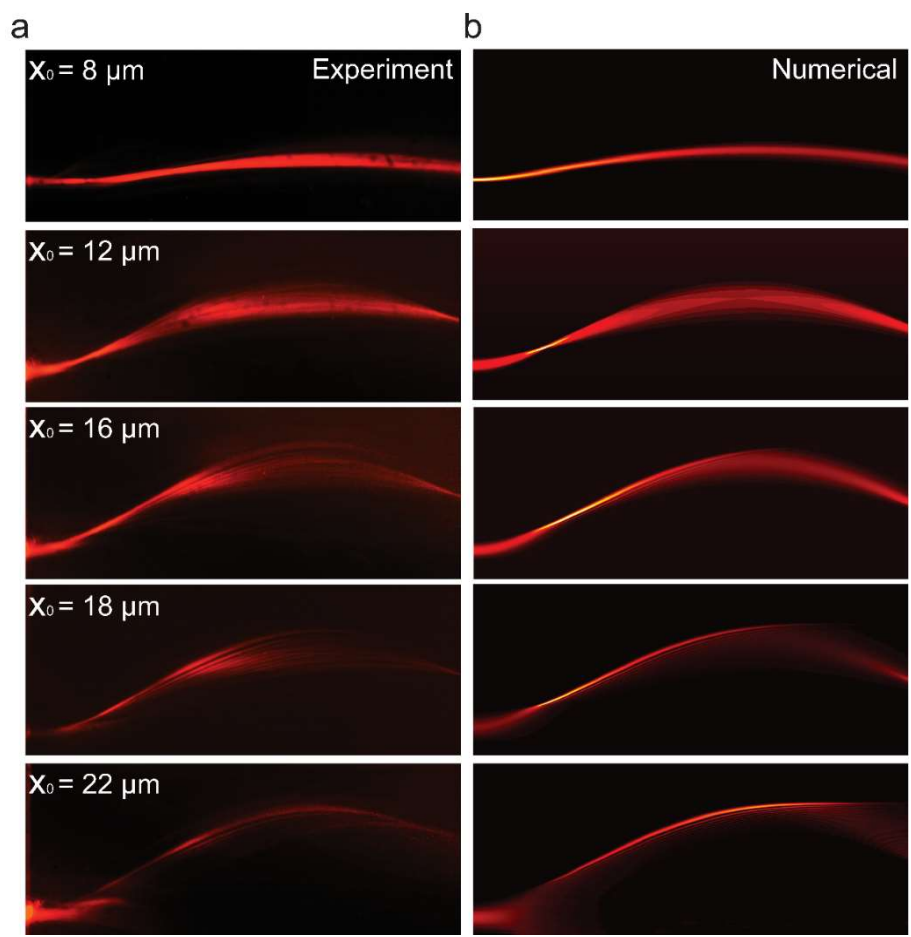

**Figure 2**. Optical measurements of the sample. (a) Field intensity observed in the experiment. The light beam bounces in the mimicked gravitational field and shows the Airy-shaped intensity profiles in the cross section. We define this as the excitation position. From top to bottom, the excitation position is, in order, 8, 12, 16, 18, and 22 μm. (b) Numerically calculated scattered field intensity obtained by full-wave continuum simulations, using the beam-propagation method. The parameters used in the simulations are extracted from the experimental data.

## 3. Optical sample measurements and simulation

A continuous-wave (cw) laser with a wavelength of 460 nm was coupled into the waveguide through a grating. When the coupled light propagates within the waveguide, it excites the rare-earth ions to emit fluorescence at 610 nm. The fluorescence emission was then collected by a microscope objective lens (Zeiss Epiplan 50×/HD 0.17, Carl Zeiss Microscopy LLC, USA) and delivered to a charge-coupled-device camera. In the experimental process, the location of the excitation point was gradually moved along the grating in one direction. Figure 2(a) displays the fluorescence pattern with the changed excitation position $x_0$. We observe that the light beams have some common features; specifically, they show oscillation in the $x$ direction during propagation, and broaden at first and then contract. However, they also show some interesting phenomena, namely that the beam cross-section displays the Airy-like intensity profile under some excitation positions, such as $x_0$=16, 18, and 22 μm. We performed the full-wave continuum simulations using the BPM, and the theoretically obtained results (figure 2(b)) are nearly identical to the experimental data. In addition, the Airy-like intensity profile is also observed.

One particular example, with an excitation position $x_0 = 18$ μm, is presented in figure 3(a); Figure 3(b) shows the corresponding simulation results. We extract the cross section intensity profiles at various propagation positions along the $z$ direction. Figure 3(c) depicts these intensity profiles from experiment (in red) and simulation (in black). The experiment and simulation results are in good agreement with each other, and a remarkable feature of the light beam evolution can be clearly seen from figure 3(c). Soon after the Gaussian beam is launched in the waveguide, the beam starts to accelerate under the external force. During its acceleration in the mimicked gravitational field, the beam width is broadened and the beam profile is changed at the same time. It can be seen that the beam is slowly transformed from the original Gaussian beam to an Airy-like beam. When the envelope reaches its maximum distance, a clear Airy-like beam can be observed in the mimicked gravitational field. As the beam propagates further, it will be transformed back into a Gaussian beam. The envelope of the beam gradually narrows and the Airy-like beam is transformed back into Gaussian beam again. The reason is that the refractive index of our sample in part of region ($-30 \leq x < 0$, figure 1 (e)) shows the linear dependence function corresponding to the uniform gravitational field based on TO works [1, 12, 15], and has the Airy function solution based on Landau's method. While the refractive index in other region ($-60 \leq x < -30$, figure 1 (e)) shows a slight deviation from the linear dependence function. Due to this slight deviation, the solution of Landau's method is slowly changing with the increasing propagation distance because different location has different potential field. After the slow and complicated change, the beam is transformed back to a Gaussian one. At this point, we have given a qualitative explanation, and what physically happens to the light beam in the mimicked gravitational field and will be discussed in the next section.

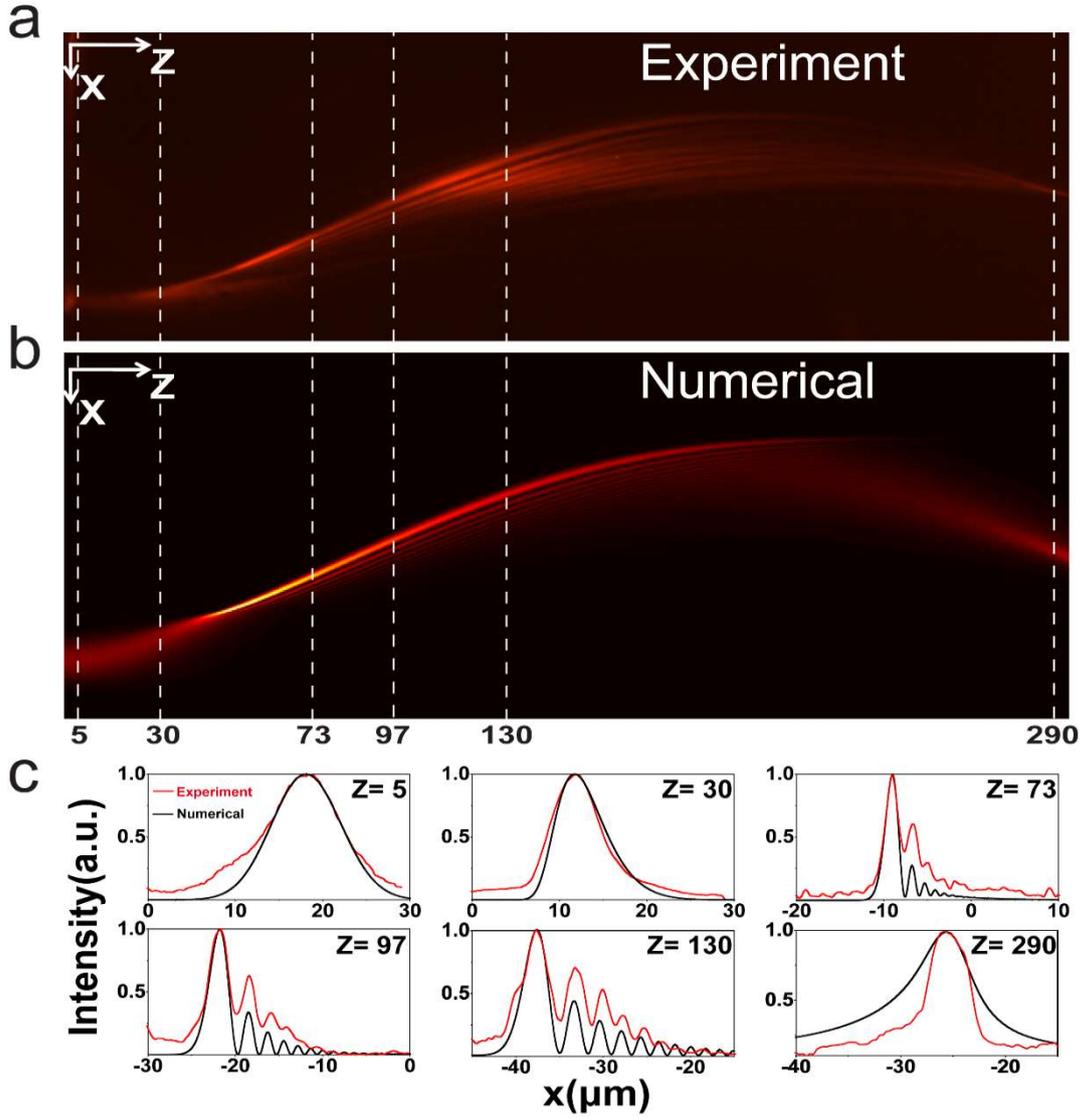

**Figure 3**. Plot of the cross section intensity profiles at distances of 4, 30, 73, 97, 130, and 290 μm along the propagation direction $z$. (a) Particular experimental example with an excitation position of 18 μm. (b) Numerically calculated beam propagation in the system with the same parameters as in (a). (c) Cross section intensity of the light beam from experiment (in red) and simulation (in black) at various positions along the $z$ direction.

## 4. Airy-shaped intensity profiles in the mimicked gravitational field

In Landau [35], a model was proposed to describe the evolution of wave packets in a gravitational field (i.e., the permittivity of the medium is a linear variation). The intensity of the light wave resembles that of an Airy function, so Landau's model can also be used in this work. To explain the Airy-shaped intensity profile in the cross section in figure 3(a), we use the reference frame shown in figure 1(c), and the gravitational field region we refer to is $x \leq 0$. We begin with the paraxial approximation to Maxwell's equations, which is mathematically equivalent to Schrödinger equation [37]:

$$i\frac{\partial \psi(x,z)}{\partial z} = -\frac{1}{2n_0 k_0}\frac{\partial^2 \psi(x,z)}{\partial x^2} - \frac{(n_{eff}^2(x)-n_0^2)k_0}{2n_0}\psi(x,z), \qquad (1)$$

where $\psi(x,z)$ is the slowly varying electric field amplitude, $x$ the transverse coordinate, $z$ the propagation distance along the waveguide axis, $k_0$ the free-space wave vector, $n_0$ the refractive index of the excitation position, and $n_{eff}^2(x)$ the effective refractive permittivity profile of the waveguide. If we focus on the stationary solutions of the Schrödinger equation, $\psi(x,z) \equiv \varphi(x)$ and $n_{eff}^2(x)$ show a linear dependence on the $x$ axis in the small range, namely $n_{eff}^2(x) = ax+b$, then Eq. (1) can be rewritten as:

$$\frac{d^2\varphi(x)}{dx^2} + (n_{eff}^2(x)-n_0^2)k_0^2 \varphi(x) = 0. \qquad (2)$$

If we define a new variable $\xi = -(x+\frac{b-n_0^2}{a})$ instead of $x$, Eq. (2) becomes

$$\frac{d^2\varphi(\xi)}{d\xi^2} - (ak_0^2)\xi\varphi(\xi) = 0. \qquad (3)$$

It is easy to find that the solution of Eq. (3) is: $\varphi(\xi) = \dfrac{\text{Ai}(\alpha^{1/3}\xi)C}{\alpha^{1/6}}$, where $\text{Ai}(x)$ is the Airy function, $\alpha = ak_0^2$, and $C$ is a constant.

We then obtain:

$$\varphi(x) = \frac{\text{Ai}(-\alpha^{1/3}(x+\beta))C}{\alpha^{1/6}}, \qquad (4)$$

where $\beta = \dfrac{b-n_0^2}{a}$.

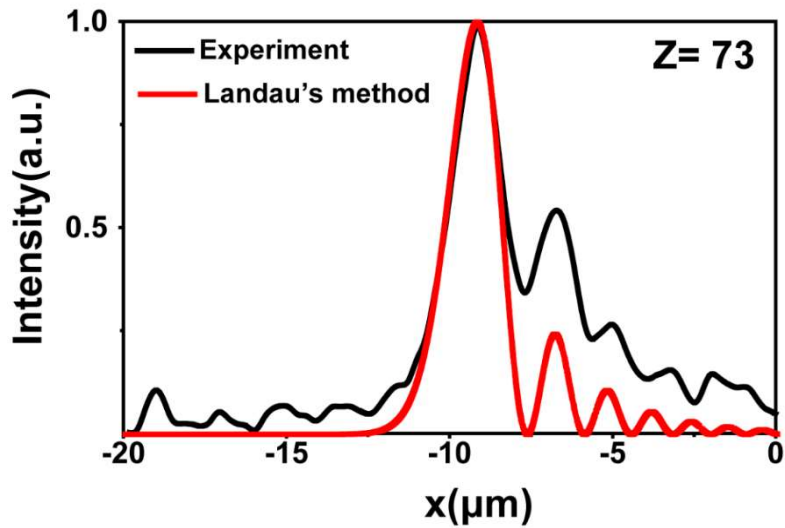

**Figure 4**. Cross section intensity profiles measured by experiment (in black) and calculation based on Landau's method (in red).

Figure 4 depicts the cross section profiles measured by experiment (in black) and calculation (in red) based on Eq. (4). We can see that the result of the theoretical calculation agrees well with that of the experiment, indicating that our experimental method can generate accelerated beams as proposed by Landau.

## 5. Conclusions

We have realized a reversible wavefront shaping technique that has not, to our knowledge, been reported before. The "gravitational field" effect is achieved using an inhomogeneous effective refractive index obtained in a TO waveguide. The evolution of a light beam in the "gravitational field" is directly observed using the fluorescence imaging method. The experimental results are confirmed with numerical simulations. Moreover, the reversible wavefront shaping process can be nicely explained by a theoretical model based on Landau's method. Such an efficient, reversible front shaping method may also be applied to generate and control light beam propagation in integrated optoelectronic elements, in addition to having more general applications in the development of new kinds of photonic devices.

## 6. Acknowledgements

This work was financially supported by the National Key Projects for Basic Researches of China (Grant Nos. 2017YFA0205700 and 2017YFA0303700) and by the National Natural Science Foundation of China (Grant Nos. 11690033, 11621091, 61425018, 11374151). H. L. thanks Rivka Bekenstein and Mordechai Segev for their helpful discussions.